\documentclass[12pt]{article}
\usepackage{amssymb,amsmath,cite} 
\catcode`\@=11
\@addtoreset{equation}{section}
\def\theequation{\arabic{section}.\arabic{equation}}
     
\catcode`\@=11
\def\thesection{\arabic{section}.}

\def\appendix{\setcounter{section}{0}
        \def\thesection{Appendix.}
        \def\theequation{\Alph{section}.\arabic{equation}}}
\def\section{\@startsection{section}{1}{\z@}{3.5ex plus 1ex minus
   .2ex}{2.3ex plus .2ex}{\large\bf}}

\def\eqnarray{\let\@currentlabel=\theequation\refstepcounter{equation}
    \global\@eqnswtrue
    \global\@eqcnt\z@\tabskip\@centering\let\\=\@eqncr
    $$\halign to \displaywidth\bgroup\@eqnsel\hskip\@centering
      $\displaystyle\tabskip\z@{##}$&\global\@eqcnt\@ne 
       \hfil${{}##{}}$\hfil
      &\global\@eqcnt\tw@ $\displaystyle\tabskip\z@{##}$\hfil 
       \tabskip\@centering&\llap{##}\tabskip\z@\cr}
\def\lefteqn#1{\hbox to 4\arraycolsep{$\displaystyle #1$\hss}}
   
\long\def\@makefntext#1{\parindent 0cm\noindent
\hbox to 1em{\hss$^{\@thefnmark}$}#1}

\newcommand{\bea}{\begin{eqnarray}}
\newcommand{\eea}{\end{eqnarray}}
\newcommand{\beq}{\begin{equation}}
\newcommand{\eeq}{\end{equation}}
\newcommand{\non}{\nonumber}
\newcommand{\lb}{\label}
\begin{document}
 \begin{titlepage}
\flushright{UCD-03-05\\
hep-th/0305113\\
May 2003}\\
\vspace*{.5cm}
\begin{center}
{\Large\bf  Quasinormal Modes\\[.5ex]
and Black Hole Quantum Mechanics\\[1ex]  
in 2+1 Dimensions}\\[2ex]

D{\sc anny} B{\sc irmingham}\footnote{birmingham@dirac.ucdavis.edu; 
on leave from Department of Mathematical Physics, University College Dublin,
Ireland}, S.\ C{\sc arlip}\footnote{carlip@dirac.ucdavis.edu}, 
Y.\ C{\sc hen}\footnote{chen@physics.ucdavis.edu}\\
\vspace{.1cm}
       {\small\it Department of Physics}\\
       {\small\it University of California}\\
       {\small\it Davis, CA 95616, USA}
\end{center}
\vspace*{1cm}
\begin{center}
{\large\bf Abstract}
\end{center}
\begin{center}
\begin{minipage}{5in}
{\small 
\noindent We explore the relationship between classical quasinormal
mode frequencies and black hole quantum mechanics in 2+1 dimensions.  
Following a suggestion of Hod, we identify the real part of the quasinormal 
frequencies with the fundamental quanta of black hole mass and angular 
momentum. We find that this identification leads to the correct quantum 
behavior of the asymptotic symmetry algebra, and thus of the dual conformal
field theory.  Finally, we suggest a further connection between quasinormal 
mode frequencies and the spectrum of a set of nearly degenerate ground states 
whose multiplicity may be responsible for the Bekenstein-Hawking entropy.
\vspace*{.25cm}
}
\end{minipage}
\end{center} 
\end{titlepage}

\section{Introduction}

Since the earliest days of black hole thermodynamics, the belief has 
been widely held that black hole quantum mechanics would one 
day lead to deep insights into quantum gravity.  The form of the 
Bekenstein-Hawking entropy strongly suggests that black holes have 
a discrete area spectrum \cite{Bekenstein}, hinting in turn at the 
quantization of fundamental geometric quantities.  This idea received 
new impetus from a suggestion of Hod \cite{Hod}, following an earlier 
qualitative observation of Bekenstein \cite{Bekenstein2}, that the classical 
``ringing'' modes of a black hole---the quasinormal modes---might 
be related by way of the correspondence principle to the elementary 
quanta of mass and angular momentum.  This suggestion received 
dramatic support from Dreyer's discovery \cite{Dreyer} that such a 
correspondence correctly fixes the Immirzi parameter \cite{Immirzi}, 
an undetermined prefactor in the area operator of loop quantum gravity.

Hod's starting point was the observation---originally numerical, but 
later confirmed analytically by Motl \cite{Motl,Motl2}---that in the large 
damping limit, the classical quasinormal modes of a Schwarzschild 
black hole of mass $M$ have a real part
\bea
\omega_{\hbox{\scriptsize QNM}} \sim \frac{\ln 3}{8\pi G M} .
\label{a1}
\eea
Since the horizon area of a Schwarzschild black hole is $16\pi G^2M^2$, 
an identification of $\hbar\omega_{\hbox{\scriptsize QNM}}$ with the 
elementary quantum of mass $\Delta M$ gives an area spacing of
$\Delta A = 4\ln3\, \hbar G$, a spectrum of a form suggested earlier for 
rather different reasons \cite{Mukhanov,BekMuk,Kastrup}.  Dreyer 
subsequently adapted this argument to loop quantum gravity.  In that 
approach to quantization, areas are not equally spaced, but one can 
still define an ``elementary'' transition $\Delta A$, corresponding to a 
change $\Delta j =\pm1$ of a state represented by a spin network 
(see \cite{Corichi}).  Remarkably, Dreyer found that the identification 
$\Delta M = \hbar\omega_{\hbox{\scriptsize QNM}}$ fixed the Immirzi 
parameter to be
\bea
\gamma = \frac{\ln3}{2\pi\sqrt2} .
\label{a2}
\eea
But the entropy of a black hole in loop quantum gravity is inversely
proportional to $\gamma$ \cite{Ashtekar}, and (\ref{a2}) is precisely 
the right value to give the correct Bekenstein-Hawking entropy.

The value (\ref{a2}) is odd enough that it is hard to attribute its appearance 
to coincidence.  But tests of the universality of this approach---looking, 
for example, at black holes in more than four dimensions, at de Sitter and 
anti-de Sitter space, and at charged and rotating black holes \cite{Motl,%
Motl2,Kunstatter,Berti,Hod2,Cardoso,Abdalla,Cardoso2,Berti2}---have 
so far been, at best, inconclusive.  In particular, quasinormal mode
boundary conditions depend not only on the geometry near the black
hole horizon, but also on the asymptotic geometry.  This can cause
quasinormal modes to behave very differently in asymptotically flat and 
asymptotically anti-de Sitter spacetimes.  On the other hand, one might
expect black hole energy levels to depend only on local physics near the
horizon, and it is not obvious how to reconcile such locality with the
quasinormal mode behavior.  We therefore decided to consider the 
(2+1)-dimensional asymptotically anti-de Sitter black hole of Ba{\~n}ados, 
Teitelboim, and Zanelli (BTZ) \cite{BTZ} to look for connections between 
classical quasinormal modes and quantization.   

The BTZ black hole differs in important ways from black holes in more 
than three spacetime dimensions (see \cite{Carlip} for a review).  In
particular, it seems likely that the horizon length, the lower dimensional
analog of area, is not quantized \cite{Freidel}.  On the other hand, a good
deal is known about the quantum mechanics of the BTZ black hole, and
the quasinormal modes are known exactly \cite{Birm,BSS1}.  If Hod's 
proposed correspondence between quasinormal modes and black hole
quantization is correct, some version should appear in this setting as well.

We find that there is, indeed, a relation between quasinormal mode 
frequencies and black hole quantization in 2+1 dimensions, but also that 
it is quite different from the Schwarzschild case considered by Hod and
Dreyer.  As we explain below, the quantum mechanics of the BTZ black 
hole is characterized by a Virasoro algebra---that is, a two-dimensional
conformal algebra---at infinity, which characterizes the dual conformal 
theory in the AdS/CFT correspondence \cite{BH,Strominger}.  Identifying 
the real part of the quasinormal frequencies with the fundamental quanta 
of black hole mass and angular momentum, we find that an elementary
excitation corresponds exactly to a correctly quantized shift of the Virasoro
generator $L_0$ or ${\bar L}_0$ in this algebra.  We also argue, somewhat
more speculatively, that the quasinormal frequencies may also be associated
with the spectrum of a collection of nearly degenerate ground states whose 
multiplicity may be responsible for the BTZ black hole entropy.

\section{Quasinormal Modes and Mass Spectrum}

The BTZ black hole is a solution of the vacuum Einstein equations in three
spacetime dimensions with a negative cosmological constant $\Lambda=
-1/l^2$.  A rotating black hole is parametrized by its ADM mass $M$ and 
angular momentum $J$. In terms of the location of the inner and outer horizons
$r_{\pm}$, we have  
\bea
M = \frac{r_{+}^{2} + r_{-}^{2}}{8Gl^{2}},\quad  J = \frac{r_{+}r_{-}}{4Gl}.
\lb{mass}
\eea
The BTZ metric is asymptotically anti-de Sitter, and as Brown and Henneaux
first showed \cite{BH}, the diffeomorphisms that preserve the asymptotic 
structure generate a left- and right-moving Virasoro algebra, with 
\cite{Strominger,Banados}
\bea
L_{0} = \frac{1}{2}(Ml + J) + \frac{l}{16G},\quad 
  {\bar L}_{0} = \frac{1}{2}(Ml - J)+ \frac{l}{16G} .
\lb{L0}
\eea
This algebra can be recognized as a two-dimensional conformal algebra, and 
the BTZ black hole provides one of the earliest and most explicit examples of
the AdS/CFT correspondence.

A detailed study of the quasinormal mode spectrum for the rotating BTZ 
black hole has been given in \cite{Birm,BSS1}, following earlier work in
\cite{Mann,Gov,Lemos}.  For a perturbation by a field with conformal weight
$(h_{L},h_{R})$, the quasinormal modes take the general form
\bea
R(r)\exp\{i(\hbox{Re}\,\omega) (t\pm l\phi)\}
  \exp\{(\hbox{Im}\,\omega) t \}
\label{mode}
\eea
with frequencies that can be written naturally in terms of the left and right 
temperatures as  
\bea
\omega_{L} &=& \frac{k_{L}}{l}- 4\pi i\; T_{L}(n + h_{L}), \non\\
\omega_{R} &=& \frac{k_{R}}{l} -4\pi i\; T_{R}(n + h_{R}),
\lb{qnm}
\eea
with a mode number $n \in {\bf N}$ and $k_{L}, k_{R} \in {\bf Z}$.
The left and right temperatures   
$T_{L,R} = (r_{+} \mp r_{-})/2 \pi l^{2}$ are related to the  Hawking 
temperature $T_H$ by  
\bea
\frac{1}{T_{L}} + \frac{1}{T_{R}} = \frac{2}{T_{H}}.
\lb{TH}
\eea

We can now apply Hod's argument to the BTZ black hole.  We first note 
that the real part of the quasinormal frequencies depends only on $k_{L,R}$
and $l$, and in particular is independent of the mode number $n$.  Thus, 
in contrast to the (3+1)-dimensional case, there is no need to go to the large 
damping limit.  Since by (\ref{mode}) the quasinormal modes carry angular 
momentum as well as energy, we require that
\bea
\Delta M &=& \omega_{L} + \omega_{R} = \frac{k_{L}}{l} +
\frac{k_{R}}{l},\non\\
\Delta \left(\frac{J}{l}\right) &=&  \omega_{L} - \omega_{R} = 
\frac{k_{L}}{l} - \frac{k_{R}}{l} .
\lb{DeltaM}
\eea
Using (\ref{L0}), we see that we are led directly to a quantization
of the Virasoro operators, with
\bea
\Delta L_{0} = k_{L},\quad\Delta {\bar L}_{0} = k_{R}.
\lb{DeltaL0}
\eea

This result can be viewed both as corroboration of Hod's correspondence
principle and as a new hint about the quantization of the BTZ black hole.  
Strominger's asymptotic symmetry analysis \cite{Strominger,BSS} does not
in itself tell us what conformal field theory describes the quantum black 
hole, and there is still a great deal of uncertainty about this issue \cite{Carlip2}.  
There are, however, strong arguments that, at least for pure gravity, the 
relevant theory should be related to an $\hbox{SL}(2,{\bf R})\times
\hbox{SL}(2,{\bf R})$ WZW model \cite{Carlip2}, probably reduced to a
Liouville theory \cite{Coussaert}.  Theories of this class may have ``ground
states'' whose values of $(L_0,{\bar L}_0)$ do not differ by integers, but
each such ground state lies at the base of a tower of states with integrally
spaced $(L_0,{\bar L}_0)$.  For the WZW model, for instance, these states
are built up by acting on a ground state with currents $J^a_{-n}$.  The
quantization (\ref{DeltaL0}) is thus exactly what one would expect for the
transitions between such states.

\section{Ground States and Liouville Theory} 

We argued above that the BTZ quasinormal frequencies give the correct
quantization for transitions among states built on a single ground state.
It is well known, however, that in Liouville theory such states are not 
sufficient to account for the BTZ black hole entropy \cite{Martinec}.  So
it is interesting to ask whether quasinormal modes can tell us anything
about possible degenerate or nearly degenerate ground states.  Although
the answer is more speculative, it is possible that they can.

We begin with a slight detour.  A point particle in (2+1)-dimensional
anti-de Sitter space can be described as a conical singularity, with deficit
angle $2\beta$ and ``time jump'' $2\pi A$.  Such a solution has an ADM 
mass lying between the anti-de Sitter value of $-1/8G$ and the extremal 
BTZ black hole value of zero \cite{BH}.  There is, however, another 
definition of the mass and angular momentum 
of a conical singularity \cite{DJH,DJ}, coming from the fact that such a 
spacetime also solves the Einstein field equations with a delta function 
stress-energy tensor.  In terms of these quantities, the deficit angle is 
$2\beta=8\pi Gm$ and the ``time jump'' is $2 \pi A=8\pi Gj$.  Since a 
point particle, like a black hole, is asymptotically anti-de Sitter, it is 
characterized by Virasoro charges \cite{BH}:  
\bea
L_{0} &=& - \frac{l}{16 G}
\left[ 1 - 4G\left(m + \frac{j}{l}\right)\right]^{2} + \frac{l}{16G},\non\\
{\bar L}_{0} &=&  - \frac{l}{16 G}
\left[ 1 - 4G\left(m - \frac{j}{l}\right)\right]^{2} + \frac{l}{16G} ,
\label{Lpt}
\eea
where we have used the conventions of \cite{Banados}.
Applied to $m$ and $j$, Hod's correspondence principle would give
\bea
\Delta\left(m+\frac{j}{l}\right) = \frac{2k_L}{l} , \quad
\Delta\left(m-\frac{j}{l}\right) = \frac{2k_R}{l} ,
\label{Deltam}
\eea
again relating left- and right-moving quasinormal modes to the holomorphic
and antiholomorphic generators of the Virasoro algebra.

Equation (\ref{Deltam}) does not yet have any obvious relationship to the
quantum mechanics of black holes.  Remarkably, though, there are a set
of states in Liouville theory with exactly the same range of conformal weights 
(\ref{Lpt}) as those of point particles.  These states---which are picked out 
naturally by the quantum group structure of Liouville theory---lie in the 
``nonnormalizable sector'' \cite{Seiberg}, the sector that is known to be needed 
to give the right counting of states for black hole entropy.  While the primary 
states in this sector have negative norm, recent work indicates that they
have descendant states of positive norm that decouple from the negative-norm
states in much the same way that the null states decouple in minimal models
\cite{Chen}.  These states thus provide a natural candidate for a set of nearly
degenerate ``ground states'' on which to build the excited states of the
preceding section.

The ``decoupled'' positive norm states in question have conformal weights 
\bea
L_{0} &=& - \frac{1}{2 \gamma^{2}}
(1 - \alpha \gamma)^{2} + \frac{1}{2 \gamma^{2}}, \non\\
{\bar L}_{0} &=& - \frac{1}{2\gamma^{2}}
(1 - {\bar\alpha} \gamma)^{2} + \frac{1}{2 \gamma^{2}},
\label{LiouL}
\eea
where $\gamma^{2} = 8G/l$ up to small quantum corrections.  Identifying
(\ref{LiouL}) with (\ref{Lpt}), we see that the transitions (\ref{Deltam}) 
correspond to
\bea
\Delta \left(\frac{\alpha}{\gamma}\right) = k_{L},\quad
\Delta \left(\frac{{\bar\alpha}}{\gamma}\right) = k_{R}.
\eea
This condition picks out a certain natural set of quantum group representations,
and calculations \cite{Chen} indicate that these representations, combined 
with their towers of excited states, give exactly the right statistical mechanical 
counting of the BTZ black hole entropy.

\section{Discussion}

We began with the question of whether Hod's suggested correspondence
principle, relating classical quasinormal mode frequencies to black hole
quantization, could be tested in three spacetime dimensions.  In one sense,
we have found that it could, and that the quantum mechanics of the BTZ 
black hole gives support to Hod's proposal.  One part of the test is fairly
clear: quasinormal mode energies and angular momenta describe transitions 
among excited states that have a clear interpretation in terms of the Virasoro 
 algebra of the dual conformal field theory.  A second piece is more tentative, 
but potentially even more exciting: the quasinormal mode correspondence
may offer us insight into the detailed construction of black hole states in 
Liouville theory.

On the other hand, our results are also rather puzzling.  Although we have
found a connection between quasinormal modes and quantum black holes
in 2+1 dimensions, the connection is very different from the one that has been 
suggested for asymptotically flat black holes in 3+1 dimensions.  We have 
not found a quantization of horizon size, and have no analog of the Immirzi 
parameter (\ref{a2}).  Given the contrast between quasinormal modes in 
asymptotically flat and asymptotically AdS spaces, differences are perhaps 
to be expected, but it is then surprising that a strong connection to 
quantization survives.  In short, while we find evidence for a correspondence 
of the sort suggested by Hod, the details of this correspondence are apparently 
considerably more complicated than has been appreciated.  We may learn
more by studying the large class of string theory black holes whose
near-horizon geometry contains the BTZ solution.  It will be interesting
to see whether the quasinormal mode structure of these higher-dimensional
black holes is consistent with the quantization we have found in 2+1 dimensions.
 
\vspace*{2ex}
\noindent{\large\bf Acknowledgments}\\ \\
D.B.\ is grateful to the Departments of Physics at U.C.\ Davis,
U.C.\ Santa Barbara, and University of the Pacific for hospitality, 
and was supported in part by Enterprise Ireland grant
IC/2002/021.  S.C.\ and Y.C.\ were supported in part by U.S.\
Department of Energy grant DE-FG03-91ER40674.

\end{document}